\documentclass{cimento}
\usepackage{graphicx}
\title{The MAGIC Telescope and the Observation of Gamma Ray Bursts}
\shortauthor{D.~Bastieri \emph{et al.}}
\author{\noindent%
  D.\ $\!$Bastieri\from{padova}%
    \thanks{Corresponding author: \emph{bastieri@pd.infn.it}.},
  N.\ $\!$Galante\from{pisa},
  M.\ $\!$Gaug\from{bcn},
  M.\ $\!$Garczarczyk\from{muc},
  F.\ $\!$Longo\from{trieste},
  S.\ $\!$Mizobuchi\from{muc}\from{ehime}
 \atque
  L.\ $\!$Peruzzo\from{padova} for the MAGIC Collaboration%
    \thanks{Updated collaborator list at
      \emph{ http://magic.mppmu.mpg.de/collaboration/index.html }}.}
\instlist{
  \inst{padova} Universit\`a and INFN -- Padova,
                Via Marzolo 8, 35131 Padova -- Italy.
  \inst{pisa} Universit\`a di Siena and INFN -- Pisa,
                Via Buonarroti 2, 56127 Pisa -- Italy.
  \inst{bcn} IFAE, Edifici Cn.\ Facultat Ci\`encies UAB,
             08193 Bellaterra (Barcelona) -- Spain.
  \inst{muc} Max Planck Institut f\"ur Physik,
             F\"ohringer Ring 6, 80805 M\"unchen -- Germany.
  \inst{trieste} Universit\`a and INFN -- Trieste,
                 Via Valerio 2, 34100 Trieste -- Italy.
  \inst{ehime} Ehime University 3,
               Bunkyo-cho, Matsuyama 790-8577 -- Japan.
}
\PACSes{\PACSit{95.55.Ka}%
               {X- and gamma-ray telescopes and instrumentation}
\PACSit{95.85.Pw}{Astronomical observations: gamma-ray}
\PACSit{98.70.Rz}{gamma-ray sources; gamma-ray bursts}}

\begin{document}

\maketitle

\begin{abstract}
The MAGIC Telescope, now taking data with an energy threshold well
below 100 GeV, will soon be able to take full advantage of the fast
slewing capability of its altazimuthal mount.  Exploiting the link
with the GCN network, the MAGIC Telescope could be one of the first
ground-based experiments able to see the prompt emission of Gamma
Ray Bursts in the few tens of GeV region.
\end{abstract}

\section{Introduction}
\noindent The MAGIC Collaboration started to build the Telescope
since the beginning of 2001.  The MAGIC Telescope entered the
commissioning phase in October 2003 and had, since then, observed
few different sources at different redshifts.

The telescope is one of the so-called \emph{last generation\/}
Cherenkov telescopes and can explore the electromagnetic spectrum
starting well below $\sim 100\,\mathrm{GeV}$.

Besides MAGIC\cite{MAGIC}, built at $2,200\,\mathrm{m}$ a.s.l.\
in the island of La Palma, Canary Islands, there are three other
Collaborations working with \emph{last generation\/} Cherenkov
telescopes: HESS\cite{HESS} (Windhoek, Namibia), CANGAROO%
\cite{CANGAROO} (Woomera, Australia) and VERITAS\cite{VERITAS}
(Horseshoe Canyon, Arizona, USA).  While these last three are
actually an array of telescopes (4 CANGAROO, 7 VERITAS and 4,
at the beginning, also HESS), MAGIC favoured the construction
of a single, although huge, telescope, featuring
$241\,\mathrm{m}^2$ of reflecting surface instead of
$\sim 100\,\mathrm{m}^2$ as the other detectors do.

Nevertheless, a second telescope, clone of the first one, with the
same dimensions and improven capabilities, is already foreseen for
MAGIC and should be operating in 2007, by the time of the launch
of the GLAST mission.

The keystones of the single MAGIC are the huge reflecting area,
capable to collect even the few Cherenkov photons emitted by low
energy gamma, and the trigger, able to cope with the high rate of
the Night Sky Background, filtering it and let the data acquisition
mostly deal with real atmospheric showers.  In this way, MAGIC
should reach the lowest energy threshold among the Cherenkov
experiments.

Additionally, the overall mechanical design of the telescope,
based upon a carbon-fiber \emph{filigree}, allows
the telescope to slew to a new position in less than 20 seconds,
in order to follow up the GRB since its early onset.

MAGIC it is not designed to give quick alerts of GRBs, mainly
because of its limited \emph{field of view\/} of $3.5^\circ$ in
diameter.  Nevertheless, it can react very quickly to alert
coming from other detectors.  In particular, MAGIC is part of the
GCN, the \emph{GRB Coordinates Network}\cite{GCN}, made up
mainly of satellites like SWIFT, that can recognise GRB events
with its on-board detectors and can quickly react sending the
celestial coordinates of the presumed GRB event.  GCN takes then
care of broadcasting the data to ground instruments such as
many optical telescopes and MAGIC.

In this work, after a brief description of MAGIC first
observations, we provide some quantitative predictions on GRB
observation for MAGIC, but most of all we show that MAGIC is
working, and due to its low energy threshold and fast repositioning
capabilities can be the ideal detector to follow the early
development of GRB at high energies.

\section{The MAGIC Telescope}
\noindent MAGIC has been in a commissioning phase since October 2003.
Beyond technical runs, there was time also for some physics
observations.  Both kind of runs were necessary to tune or improve
the hardware of the detectors, and the first physics signals were
ready in February 2004.

The first sources observed were the Crab Nebula and \textsc{Mkn~421}.
Both of these sources are a sort of \emph{standard\/} sources: the
Crab has a steady flux that can be used to calibrate detectors, and
\textsc{Mkn~421}, while not being steady, stays from time to time
in a high state with fluxes few times bigger than the Crab one.

These two sources were observed in Winter 2003/2004 (the Crab Nebula)
and during Spring 2004 (\textsc{Mkn~421}) and both revealed signals
well above the $5\sigma$ level.

\begin{figure}
\noindent\includegraphics[width=\textwidth]{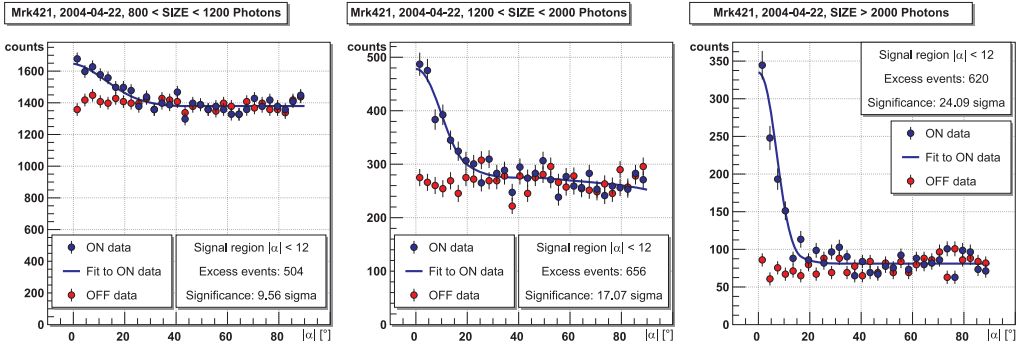}
\caption{\label{fig:mkn421}Data from 155-minutes of acquisition
  taken on April $22^{\mathrm{nd}}$, 2004 with MAGIC pointing at
  \textsc{Mkn~421}.  The histogram of \emph{alpha\/} for on- and
  off-source observations, are relative to different bin in
  \emph{size}: 800--1200 (left), 1200--2000 (middle) and more than
  2000 photons (right).\vspace*{-.15in}}
\end{figure}
Special attention must be given to Fig.~1.  It represents
the data collected on April $22^{\mathrm{nd}}$, 2004 pointing
at \textsc{Mkn~421} that was in a high state.  The so-called
IACT (Imaging Atmospheric Cherenkov Technique) used by MAGIC,
records for each event the directions of the Cherenkov photons
emitted during the shower development.  The image is then
analysed using a well-established technique that exploits
the so-called \emph{Hillas parameters}\cite{Hillas}.

Two Hillas parameters have a particular importance in the figure:
\emph{alpha\/} and \emph{size}.  \emph{Alpha\/} is related to the
actual direction of the primary particle that initiated the shower,
thus an excess in \emph{alpha\/} must be seen in the direction of
a source. \emph{Size\/} is the number of photons making up the
image and is related to the energy of the primary particles (the
more energetic is the particle, more photons appear in the image).

The three figures show that the \textsc{Mkn~421} flare was well
detected by MAGIC at different energies, and the excess seen in
the first figure with \emph{size\/} ranging from 800 to 1200
photons is consistent with an energy well below
$100\,\mathrm{GeV}$.

\section{GRB Observability}\vspace*{-.05in}
\noindent Having shown that MAGIC is effectively working, let us 
see what we expect from GRBs.  The main parameter influencing the
GRB observability is the telescope
\emph{duty-cycle}\cite{ICRC:GRB}.  The duty-cycle of the telescope
was calculated as the fraction of sky area accessible to MAGIC per
year, where, to be accessible, the following conditions must be met:
\vspace*{-.05in}
\begin{itemize}
  \item the Sun must be below astronomical horizon or
        have a zenith angle $>108^\circ$;\vspace*{-.05in}
  \item the minimum angular distance from the Moon
        must be $30^\circ$;\vspace*{-.05in}
  \item the relative humidity of the air must be lower than 80\%;\vspace*{-.05in}
  \item the wind speed must be lower than $10\,\mathrm{m/s}$.\vspace*{-.05in}
\end{itemize}

The calculation, using the weather data and ephemerides for year
2001, gives a value slightly below 10\%.  This means that roughly
$\frac{1}{10}$ of all GRB alerts can be immediately pursued.

Once determined the fraction of GRB observable, it is possible to
estimate how many GRB can be actually observed, that is, if their
significance is above $5\sigma$.  The number of $\sigma$ can be
calculated using the Li-Ma formula:
\begin{equation}
N_\sigma=Q\frac{(R_{\mathrm{on}}-R_{\mathrm{off}})T}
               {\sqrt{(R_{\mathrm{on}}+R_{\mathrm{off}})T}}
        =Q\frac{R_\gamma\sqrt{T}}
               {\sqrt{R_{\mathrm{on}}+R_{\mathrm{off}}}}
\end{equation}
where $Q$, the so called \emph{quality factor}, represents
the efficiency of the analysis and is of order unity at low
energies, $T$ is the actual acquisition time and the $R$'s
represent the rates.  Each $R$ takes into account the particle
flux, $\Phi$, and the detector sensitivity to the flux itself,
expressed as the \emph{effective area\/} of the telescope,
$A_{\mathrm{eff}}(E,\vartheta)$.  For instance, we have that
the expected signal from the GRB is:
$R_\gamma=\int\Phi_{\mathrm{GRB}}(E)
              A^\gamma_{\mathrm{eff}}(E,\vartheta)\mathrm{d}E$
and the cosmic-ray background is:
$R_{\mathrm{off}}=\int\Phi_{\mathrm{CR}}(E,\vartheta)
       A^{\mathrm{CR}}_{\mathrm{eff}}(E,\vartheta)\mathrm{d}E$.

The effective area of a Cherenkov telescope is quite huge,
reaching $10^5\,\mathrm{m}^2$ around $100\,\mathrm{GeV}$ and
dropping to $10^4\,\mathrm{m}^2$ around $50\,\mathrm{GeV}$.

The GRB flux is obtained extrapolating the power-law spectrum
found inside the BATSE catalogues, taking also into account the
cosmological cutoff due to the absorption of high-energy $\gamma$
by the Metagalactic Radiation Field via
$\gamma_{\mathrm{HE}}\gamma_{\mathrm{MRF}}
  \rightarrow\mathrm{e}^+\mathrm{e}^-$.

Of paramount importance for a good observation of GRBs is the
delay between the GRB onset and the time when MAGIC can actually
observe the GRB.  This delay sums up three different times:%
\vspace*{-.05in}
\begin{enumerate}
  \item the time needed by the satellite to recognise that
        a GRB event may be going on;\vspace*{-.05in}
  \item the time elapsing between the sending of the alert by the
        satellite, the broadcast via GCN and the final
        acknowledgement by MAGIC ($\sim 2$ seconds);\vspace*{-.05in}
  \item the slewing time needed by MAGIC to point to the GRB
        position ($<20$ seconds).\vspace*{-.05in}
\end{enumerate}

It can be calculated that reducing this total delay from
1 minute to 15 seconds increases of a factor 5 the number
of observable GRBs.

It should be noted also that the zenith angle, $\vartheta$, enters
the formulae for the rates.  The number of observable GRBs thus it
is not obtained with a straightforward multiplication of the
duty-cycle $(\approx 10\%)$ times the number of GRBs expected to
have a significance $N_\sigma>5$, but it is better expressed with
$N_{\mathrm{GRB}}=\sum_\vartheta D(\vartheta)\eta(\vartheta)$
where $D(\vartheta)$ is the duty-cycle in a given bin of zenith
angle, $\eta(\vartheta)$ is the number of GRBs (extrapolated from
BATSE catalogues) actually observable in a given bin of zenith
angle and the sum is performed over all zenith angle bins.

Values of $\eta(\vartheta)$ range between 5 and 24, strongly
depending upon the delays from the GRB onset to the actual
observation of the GRB by MAGIC, and the final result is that
MAGIC should observe $0.5\div 2$ GRBs per year.\vspace*{-.05in}

\section{Conclusions}.\vspace*{-.05in}
\noindent The results exposed in this work show that MAGIC is now
operating and can be used to follow the GRB events since its early
onset.  Observations with Cherenkov detectors below
$100\,\mathrm{GeV}$ seem now attainable, even if the analysis of
low-energy events is still a challenge and may require more work.

\acknowledgments.\vspace*{-.05in}
\noindent The development and construction of the MAGIC Telescope
was mainly supported by the BMBF (Germany), CICYT (Spain), INFN
and MURST (Italy).  The authors wish to thank also the people from
the Observatorio del Roque de Los Muchachos for the excellent
working conditions in La Palma.

\end{document}